\documentclass[aps,prc,groupedaddress]{revtex4}

\usepackage{graphicx}
\usepackage{dcolumn}
\usepackage{bm}

\begin{document}


\title{Empirical Fit to Precision Inclusive Electron-Proton 
Cross Sections in the Resonance Region} 



\author{M.E.~Christy}
\email{christy@jlab.org}

\affiliation{Hampton University, Hampton, Virginia 23668}

\author{P.E.~Bosted}
     \email{bosted@jlab.org}
\affiliation{Jefferson Lab, Newport News, Virginia 23606}


\date{\today}

\pacs{25.30.Fj,13.60.Hb, 14.20 Gk}

\begin{abstract}
An empirical fit is described to  measurements of inclusive 
inelastic electron-proton cross sections in the kinematic range
of four-momentum transfer $0 \le Q^2<8$ GeV$^2$ and 
final state invariant mass $1.1<W<3.1$ GeV. 
The fit is constrained by 
the recent high precision longitudinal and transverse (L/T) 
separated cross section measurements from 
Jefferson Lab Hall C, un-separated Hall C measurements up to 
$Q^2$ $\approx 7.5$  ${\rm GeV}^2$, and 
photoproduction data at $Q^2 = 0$. Compared to previous fits,
the present fit covers a wider kinematic range, fits both
transverse and longitudinal cross sections, 
and features smooth transitions to the 
photoproduction data at $Q^2=0$ and DIS data at high $Q^2$ and $W$.
\end{abstract}

\pacs{}

\maketitle


\section{Introduction}

Knowledge of the inclusive electron-proton cross section in the nucleon resonance region is important 
input for many research activities in nuclear and particle physics.  Classic examples include calculations 
of radiative corrections to cross sections and extractions of spin structure functions from asymmetry 
measurements.  In addition, the latter requires knowledge of the separated longitudinal and transverse 
photo-absorption cross sections.
More recently, the electron-proton resonance region cross section has been used in constraining the vector 
coupling in models~\cite{paschos,olga} of 
low energy neutrino-nucleon cross 
sections.  This is important since the quality of low energy neutrino-nucleon cross section models will 
become one of the largest uncertainties in the extraction of neutrino oscillation parameters from 
future long-baseline experiments.

Having the best possible fit to the inclusive electron-proton cross section is important for all the endeavors 
described above.  In this paper we will describe a fit to precision measurements of inclusive electron-proton 
cross sections in the resonance region for $Q^2$~$<$~7.5~${\rm GeV}^2$.  Among the advantages 
of this fit over previous resonance region proton 
fits~\cite{stuart,ioana_thesis} is the constraint on the individual 
transverse and longitudinal cross sections afforded by the use of the recent high precision L/T separated 
cross section measurements~\cite{e94110,liang} from Jefferson 
Lab Hall C, a smooth transition to the photoproduction 
point, and use of threshold-dependent Breit-Wigner forms for 
all resonances.  It needs to be stressed that while 
the fit form utilized is, in general, physically motivated, 
the focus of the current work is to provide the 
best description of the inclusive proton data and not to 
determine the masses, widths, transition form factors, 
and branching ratios of the produced resonant states.  
These are best determined from exclusive meson production data.

\section{Inclusive electron scattering from the proton}
In terms of the incident electron energy, $E$, the 
scattered electron energy, $E^{'}$, and the scattering angle, 
$\theta$, the absolute value of the exchanged 4-momentum 
squared in electron-proton scattering  is given by
\begin{equation}
Q^2 = (-q)^2 =  4EE^{'}{\sin}^2 \frac{\theta}{2}, 
\end{equation}
and the mass of the undetected hadronic system is
\begin{equation}
W^2 = M_p^2 + 2M_p\nu -Q^2,  
\end{equation}
with $M_p$ the proton mass and $\nu = E-E^{\prime}$.  In these expressions we have neglected 
the electron mass which is negligible for the kinematics studied. 

In the one-photon exchange approximation, the spin-averaged
cross section for inclusive
electron-proton scattering can be expressed in terms of the 
photon helicity coupling as
\begin{equation}
\frac{d\sigma}{d\Omega dE^{'}} = \Gamma\left[\sigma_T(W^2,Q^2) + \epsilon \sigma_L(W^2,Q^2)\right],
\label{eq:cs1}
\end{equation}
where $\sigma_T$ ($\sigma_L$) is the cross section 
for photo-absorption of purely
transverse (longitudinal) polarized photons,
\begin{equation}
\Gamma = \frac{\alpha E^{'}(W^2 - M_p^2)}{(2 \pi)^2 
Q^2 M_p E (1 - \epsilon)}
\end{equation}
is the flux of virtual photons, and
\begin{equation}
\epsilon = \left[1 + 2(1+\frac{\nu^2}{Q^2}) 
{\tan}^2 \frac{\theta}{2}\right]^{-1}
\end{equation}
is the relative flux of longitudinal virtual photons.
Since $\Gamma$ and $\epsilon$ are purely kinematic factors,
it is convenient to define the reduced cross section 
\begin{equation}
\sigma_r = {1 \over \Gamma} \frac{d\sigma}
{d\Omega dE^{'}} = \sigma_T(W^2,Q^2) + \epsilon \sigma_L(W^2,Q^2).
\label{sig_reduced}
\end{equation}
All the hadronic structure information is, 
therefore, contained in $\sigma_T$ and $\sigma_L$, which are only dependent 
on $W^2$ and $Q^2$.  

In light of the current discrepency between the proton elastic
form factors extracted from unpolarized scattering to those extracted
from polarization observables~\cite{Arrington:2007ux},
it is worth commenting on the assumed validity of the one-photon exchange 
approximation.  While, the most commonly accepted explaination for resolving 
this discrepency is the existance of two-photon (2-$\gamma$)exchange 
contributions to the elastic scattering, the available data indicate
that the reduced cross section is still quite linear in $\epsilon$ over the 
ranges measured.  Indeed, a recent search for non-linearities in the
combined elastic and inelastic data which separated the longitudinal and 
transverse cross sections (including a significant portion of the resonance 
region and DIS data which is fit here) found no significant
(i.e. less than 1\% typically) evidence for non-linearities~\cite{nonlinear}.
Hence, although it is impossible to rule out the existence of 2-$\gamma$ 
contributions to the inelastic cross sections from this data, the existence 
of such contributions does not affect our ability to provide a reliable fit to
the {\it measured} cross section data within the one-photon approximation.  
It is only the interpretation of the separated structure functions which
could be incorrectly interpreted if significant 2-$\gamma$ contributions
exist.

\section{Description of the Fit}

In the past, fits to resonance region cross sections have
typically not included data covering a large enough range
in $\epsilon$ and/or of high enough precision to
constrain both $\sigma_T$ and $\sigma_L$.  Therefore, such fits
have relied on an educated guess for the ratio
$R = \sigma_L / \sigma_T$ in order to extract data on $\sigma_T$
from the measured reduced cross sections, and it
was the extracted $\sigma_T$ that was consequentially fit.  This
guess has typically been to use a fit to deep
inelastic scattering (DIS) data on $R$ extrapolated into the resonance
region.
The assumption here is that all the resonant structure cancels in $R$.
However, it is now clear from the recent Jefferson Lab L/T data~\cite{e94110}
that the resonant structure in $\sigma_L$  differs significantly from that in 
$\sigma_T$, resulting in $W$ and $Q^2-$ dependent variations of up to 0.1 in $R$.



Utilizing the new precision data, the reduced cross 
section was fit by parameterizing $\sigma_T(W^2,Q^2)$ 
and $\sigma_L(W^2,Q^2)$ and then minimizing the 
difference of $\sigma_r$ constructed from Equation~\ref{sig_reduced} 
with respect to the data.  
The cross section parameterization contained 75 free parameters 
which were determined from the fit: 7 for the resonance masses, 7 for the 
resonance widths,
25 to describe the $Q^2$ dependence of the transverse transistion form factors, 
18 to describe the $Q^2$ dependence of the longitudinal transistion form 
factors, 10 to describe the non-resonant contribution to $\sigma_T$, 7 to 
describe the non-resonant contribution to $\sigma_L$, and a single damping 
parameter for the delta resonance.  A detailed descripton of the parameterization 
is given later in this section.  

In order to avoid local 
minima and to speed up the convergence the starting parameters for $\sigma_T$ 
and $\sigma_L$ were determined by first fitting the approximate separated cross 
sections independently. 
To accomplish this the following procedure was utilized:  
first, the $R_{1998}$~\cite{r1998} fit to DIS data on $R$ was 
used to extract $\sigma_T$ and $\sigma_L$ from 
$\sigma_r$; second, the extracted data on $\sigma_T$ and $\sigma_L$ were 
fit independently, and finally, $R$ determined 
from the new fits was used to extract $\sigma_T$ and $\sigma_L$ for the 
next iteration.  Several iterations were 
performed to ensure that reasonable convergence was 
obtained.  The fit forms 
for the resonant and non-resonant contributions to the 
cross sections were determined from this procedure for both 
$\sigma_T$ and $\sigma_L$ and were then used in a 
single fit to $\sigma_r$ in which all transverse and longitudinal 
parameters were allowed to vary simultaneously.

The fit form used to describe the separated resonance 
region photo-absorption cross sections was based on the following 
ansatz:  
1) the cross section is the incoherent sum of contributions from 
resonance production ($\sigma^R$) and a 
non-resonant background ($\sigma^{NR}$),
2) the resonant cross section can be described by 
threshold-dependent relativistic Breit-Wigner forms with $Q^2$-dependent 
amplitudes for each resonance, 
and 3) the non-resonant background varies smoothly with $W^2$.  Therefore, 
\begin{equation}
  \sigma_{T,L} (W^2,Q^2) = \sigma^R_{T,L} + \sigma^{NR}_{T,L},   
\end{equation}
with resonant contribution 
\begin{equation}
  \sigma^R_{T,L} (W^2,Q^2) = W 
\sum_{i=1}^{7} BW^i_{T,L}(W^2) \cdot [A_{T,L}^i(Q^2)]^2.  
\end{equation}
The Breit-Wigner form utilized is 
\begin{equation}
 BW^i = {K_i K^{cm}_i \over K K^{cm}} \cdot 
{\Gamma^{tot}_i \Gamma_i^{\gamma} \over \Gamma_i 
\left [ (W^2 - M_i^2)^2 + (M_i \Gamma^{tot}_i)^2 \right ]}, 
\end{equation}
with 
\begin{equation}
K = {W^2 - M_p^2 \over 2 M_p},
\end{equation}
\begin{equation}
K^{cm} = {W^2 - M_p^2 \over 2 W},
\end{equation}
\begin{equation}
K_i = K|_{M_i},
\end{equation}
and
\begin{equation}
K^{cm}_i = K^{cm}|_{M_i}.
\end{equation}
Here, $K$ and $K^{cm}$ represent the equivalent 
photon energies in the lab and center of mass (CM) frames, respectively, 
while $K_i$ and $K^{cm}_i$ represent the same 
quantities evaluated at the mass of the $i^{th}$ 
resonance, $M_i$.  $\Gamma^{tot}_i$ is 
the full decay width defined by 
\begin{equation}
\Gamma^{tot}_i = \sum_{j=1}^{3} \beta^j_i \Gamma^j_i,
\end{equation}
with $\beta^i_j$ the branching fraction to 
the $\rm j^{th}$ decay mode for the $\rm i^{th}$ 
resonance and $\Gamma^j_i$ the partial 
width for this decay mode.  For single-meson decay 
modes the partial widths were determined from  
\begin{equation}
\Gamma^j_i = \Gamma_i \left [  { p^{cm}_j 
\over p^{cm}_j |_{M_i}} \right ]^{2l+1} \cdot 
\left [  {(p^{cm}_j|_{M_i})^2  + X_i^2 
\over  (p^{cm}_j)^2 + X_i^2} \right ]^l,
\end{equation}
where $ \Gamma_i$ are the intrinsic widths of each resonance, 
the $p^{cm}_j$ are meson momenta in the center of 
mass, $l$ is the angular momentum of the resonance, 
and $X_i$ is a damping 
parameter.  In the present analysis it was found 
that a good fit to the data was obtained using 
$X_i = 0.215$ GeV for all the resonances 
except the $P_{33}(1232)$, where a smaller value 
of $X_1 = 0.1446$ GeV resulted in a modest improvement.  
Overall the goodness of the fit 
was found to be relatively insensitive to the 
exact value of $X_i$.  For the two-pion decay mode the 
partial widths were determined from 
\begin{equation}
\Gamma^j_i = \frac {W \Gamma_i} {M_i} \cdot  
\left [  { p^{cm}_j \over p^{cm}_j |_{M_i}} \right ]^{2l+4} \cdot 
\left [  { (p^{cm}_j|_{M_i})^2  + X_i^2 \over
(p^{cm}_j)^2 + X_i^2 } \right ]^{l+2}.
\end{equation}
The virtual-photon width was defined by:
\begin{equation}
\Gamma_i^\gamma = \Gamma_i  \left [  { K^{cm} 
\over K^{cm} |_{M_i}} \right ]^2 \cdot 
\left [  {(K^{cm}|_{M_i})^2  + X_i^2 
\over  (K^{cm})^2 + X_i^2} \right ].
\end{equation}

\subsection{Constraints on the $Q^2$ dependence for $\sigma_L$}

There were several physics constraints on the $Q^2$ dependence of the cross section 
which were realized in the fit form.  For instance, the longitudinal cross 
section must vanish at the photoproduction point due to current conservation, and 
at large $Q^2$ due to helicity conservation in scattering from spin-$1/2$ fermions.  
These constraints were independently imposed in both the resonant and non-resonant 
contributions to the cross section.  This was accomplished for the resonant contribution 
by a suitable form for the longitudinal transition amplitudes, which 
vanished at $Q^2 \rightarrow$ 0 and $Q^2 \rightarrow \infty$.

\subsection{Resonances included in the fit}

Seven possible resonance contributions were included in the 
fit to $\sigma_T$, and represent those with the largest photo-couplings 
to the proton as listed by the particle data group~\cite{pdg}.      
These consisted 
of the first prominent resonance, the 
$\Delta$ $P_{33}(1232)$, two states in the second 
resonance region, the $S_{11}(1535)$ and the 
$D_{13}(1520)$, and two in the third 
resonance region, the $S_{15}(1650)$ and 
the $F_{15}(1680)$.  In addition, the Roper, 
$P_{11}(1440)$ and a broad resonance around 
$W \approx 1.9$~GeV were included.  Only the pion, 
eta, and 2-pion decay modes were included for each 
resonance and the branching 
fractions used are given 
in Table~\ref{table_res_inc}.  The region 
at $W \approx 1.9$~GeV is occupied by 
many states and was assigned an angular 
momentum of $l = 3$ in the fit.  We have not tried to be exhaustive by including 
all known resonant states, but have, rather, tried to include enough of the 
dominant states to provide for a good fit to the data without requiring a burdensome 
number of additional parameters.  For recent reviews of electromagnetic 
meson production in the nucleon resonance region we refer the reader to \cite{burkert-lee}, 
and to the latest fit to the unitary isobar model, MAID2007~\cite{maid2007}.  

 



The resonances included for $\sigma_L$ were largely the same as for $\sigma_T$.  However, 
due to the decreased sensitivity of the cross section data to $\sigma_L$ it was found that including more than 
a single resonance in the region of the $D_{13}(1520)$ and $S_{11}(1535)$ around $W \approx 1.5$ had little 
impact on the goodness of the fit.  Therefore, the second resonance listed in Table~\ref{table_res_inc}
($S_{11}(1535)$) was not included in the fit for $\sigma_L$.  The choice of excluding this particular resonance 
was arbitrary as it was found that excluding the $D_{13}(1520)$ instead had a neglible impact.


\begin{table}[tbh]
\begin{center}
\begin{tabular}{c l l l l}
\hline
\hline
$I$ \hspace{0.5cm} & State  \hspace{0.5cm} &  $\beta_{1\pi}$ \hspace{0.5cm}  
& $\beta_{2\pi}$ \hspace{0.5cm}  &  $\beta_{\eta}$ \\
\hline
    1                    &     $P_{33}(1232)$       &    1.0          &    0.0          &    0.0         \\
    2                    &     $S_{11}(1535)$       &    0.45          &    0.10         &    0.45        \\
    3                    &     $D_{13}(1520)$       &    0.65         &    0.35         &    0.0         \\
    4                    &     $F_{15}(1680)$       &    0.65         &    0.35         &    0.0         \\
    5                    &     $S_{15}(1650)$       &    0.4          &    0.5          &    0.1         \\
    6                    &     $P_{11}(1440)$       &    0.65         &    0.35         &    0.0         \\ 
    7                    &     ($l=3$ assumed)     &    0.5          &    0.5          &    0.0         \\
\hline
\hline
\end{tabular}
\label{table_res_inc}
\end{center}
\caption{Resonance number $I$, name (and quantum numbers), 
and branching fractions for the resonant states 
included in the fit.}
\label{table_res_inc}
\end{table}

\subsection{Resonance transition amplitudes}

For the transverse resonance transition amplitudes the fit 
form utilized was 
\begin{equation}
  A_T^i(Q^2)  = {A_T^i(0) \over (1 + Q^2/0.91)^{c_i} } 
\cdot \left( 1 + {a_i Q^2 \over 1 + b_i Q^2} \right ), 
\end{equation} 
while for the longitudinal resonance transition amplitudes 
the fit form utilized was 
\begin{equation}
  A_L^i(Q^2)  = A_L^i(0) \cdot {Q^2 \over (1 + d_i Q^2)} e^{-e_i Q^2}. 
\end{equation} 
For large $Q^2$, $A_T^i(Q^2)$ reduces to 
\begin{equation}
  A_T^i(Q^2)  = {A_T^i(0) \over (1 + Q^2/0.91)^{c_i}}
 \cdot {a_i \over b_i}, 
\end{equation} 
which is just the dipole form for $c_i$ = 2.
For $Q^2 \rightarrow 0$,  $A_T^i(Q^2)$ linearly approaches 
the photoproduction value, $A_T^i(0)$, while $A_L^i(Q^2)$ 
linearly vanishes.

\subsection{Non-resonant background}

For the non-resonant part of the transverse cross section 
the fit form which was found to describe the data well was 
\begin{equation}  
\sigma_T^{NR}  = x^{\prime}\sum_{i=1}^2  {\sigma_T^{NR,i}(0)  
\over (Q^2+a^T_i)^{[b^T_i+c^T_iQ^2+d^T_iQ^4]}} (\Delta W)^{i+1/2}, 
\end{equation}
where 
\begin{equation}
x^{\prime} = \left ( 1 + {W^2 -(M_p + m_{\pi})^2 \over Q^2 + Q^2_o}\right )^{-1},
\end{equation}
with $m_{\pi}$ the pion mass, $Q^2_0 = 0.05$ GeV$^2$, 
and $\Delta W = W - m_{\pi}$. 
Previous resonance region fits 
have also included terms to describe the $W$ dependence 
with $(\Delta W)^{i+1/2}$.  However, these did not attempt 
to fit beyond $W \approx 2$ GeV, where, in contrast to 
the data, this particular form continues to rise with increasing 
$W^2$.  Here it was found that this rise can be 
moderated by the additional $x^{\prime}$ factor.  This allows a 
good fit to the data to be obtained up to $W = 3.2$ GeV.  
The factor $x^{\prime} \approx x = Q^2/2M_p\nu$ for 
$W^2-M_p^2 >> M_pm_{\pi}$ and 
$Q^2 >> Q^2_0$, but has the important properties, 
required to fit the entire resonance data set,
 that it approaches a constant for $Q^2 \rightarrow 0$ 
and vanishes at pion threshold.    

For the longitudinal cross section it was found that 
the non-resonant background could {\it not} be well described by powers 
of $W^{1/2}$ at fixed $Q^2$, but was better represented 
away from pion threshold by $(1-x)^{a}$.  Therefore, the form 
utilized was
\begin{equation}  
\sigma_L^{NR}  = \sum_{i=1}^1  \sigma_L^{NR,i}(0)
 {(1-x^{\prime})^{[ a^L_i t + b^L_i]} \over 
(1-x)} {(Q^2)^{c^L_i} \over (Q^2 + Q_0^2)^{(1 + c^L_i)}}
\cdot (x^{\prime})^{[d^L_i + e^L_i t]}
\end{equation}
where 
\begin{equation}
t = {\log (\log({[Q^2 + m_0] \over 0.33^2}) \over \log({m_o \over 0.33^2})} 
\end{equation}
is a slowly varying function of $Q^2$.
For $\sigma_L^{NR}$, we used  $Q^2_0 = 0.125$ GeV$^2$ and
$m_0= 4.2802$ GeV.

\section{The data sets}
A summary table of the data sets, including the number of data 
points and the $Q^2$ range of each set, is provided 
in Table~\ref{datasets}.

The nucleon resonance region ($W<2$ GeV) $ep$ scattering data 
included in the fit came principally from a series 
of Jefferson Lab Hall C experiments which utilized 
the well-understood High Momentum Spectrometer (HMS).  The 
various Hall C data sets have been estimated to have normalization 
uncertanties of $\approx 2$\% and were found to agree in their regions 
of kinematic  overlap to better than 1\%.  There also exists a large 
body of inclusive $ep$ data from the Hall B CLAS collaboration  in 
the kinematic region fit.  However, in additon to having significantly 
larger systematic uncertainties, only results for the $F_2$ structure 
function have been published.  Therefore, this data was not utilized 
since only cross section data were fit.
    
For $0.18 < Q^2 < 4.5$ the inclusive proton data 
utilized was dominated by the recent E94-110 
high precision data~\cite{e94110}, which have typical point-to-point 
uncertainties in $\epsilon$ of less than 2\%.  This experiment was optimized 
to separate $\sigma_L$ and $\sigma_T$ from Rosenbluth separations~\cite{rosen} 
and therefore, provided a significant $\epsilon$ range at fixed $W^2$ and 
$Q^2$ for constraining the separated cross sections. For $4.5 < Q^2 < 7.5$ the 
resonance data aremainly  from experiment E00-116~\cite{e00116}.  
For this data set the uncertainties were typically statistically dominated.  In 
addition, the cross sections at each $W^2$ and $Q^2$ 
were only measured at a single kinematic setting and, hence, these data did not provide 
any $\epsilon$ range to constrain $\sigma_L$ 
and $\sigma_T$ when taken by itself.  Preliminary
data from E00-002~\cite{Edwin} were used to extend
to $Q^2$ range of precision data down to 0.05 GeV$^2$.
Data in the DIS region~\cite{whitlow} ($W>2$) were used to constrain
the fit up to $W=3.1$ GeV. Several sets of photoproduction
data were used to constrain the fit at $Q^2=0$ GeV$^2$: data 
taken prior to 1975~\cite{photo_bloom,photo_armstrong,photo_meyer} 
and the more recent data from DAPHNE~\cite{daphne-p}.  While the 
latter is much more precise, it does not extend beyond the second 
resonance region. 


\begin{table}[tbh]
\begin{center}
\begin{tabular}{l l l c}
\hline
\hline
Data Set  \hspace{0.5cm}   & $Q^2_{Min}$  \hspace{0.5cm}    &  $Q^2_{Max}$  \hspace{0.5cm}   &  \# Data Points        \\
          &  ($\rm GeV^2$)  &  ($\rm GeV^2$) &                                                      \\  
\hline    
E94-110~\cite{e94110}    &    0.18      &    5     &                      1259                \\
E00-116~\cite{e00116}   &    3.6       &   7.5    &                       256                \\
E00-002~\cite{Edwin}    &    0.06      &   2.1    &                      1346                \\
SLAC DIS~\cite{whitlow}  &    0.6       &   9.5    &                           296                \\
Photoproduction (Old)~\cite{photo_bloom,photo_armstrong,photo_meyer}     &    0          &      0         &     242               \\
Photoproduction (DAPHNE)~\cite{daphne-p}  \hspace{0.5cm} &      0        &    0      &               57           \\ 
\hline
\hline
\end{tabular}
\caption{Summary of data sets included in the fit.}
\label{datasets}
\end{center}
\end{table}

\section{Results and Discussion}


\begin{figure}
\includegraphics[width=15cm,height=14cm]{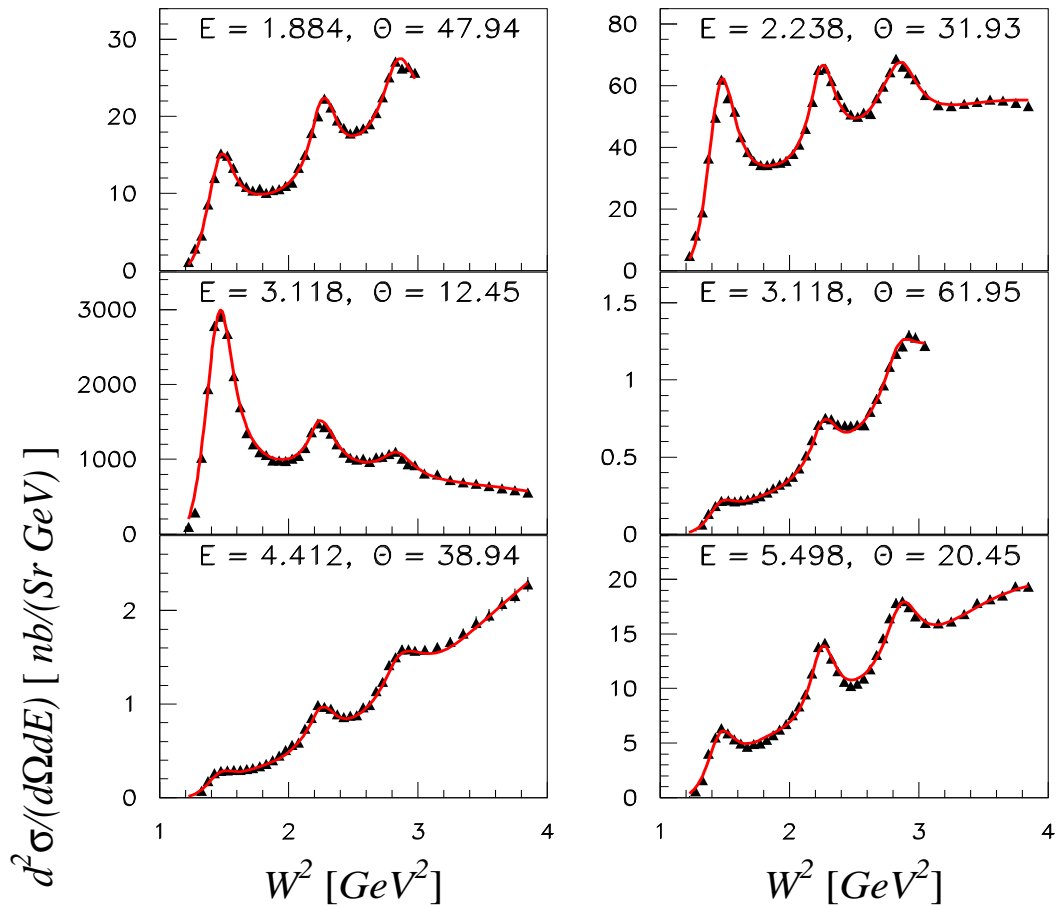}
\caption{\label{cscomp_res} (Color online) Comparison of 
the fit results (solid lines) to the JLab E94-110 resonance 
region data~\protect{\cite{e94110,liang}} (solid triangles) 
versus $W^2$, for several
representative kinematic settings. Beam energies $E$ are
in GeV, and electron scattering angles $\theta$ are in
degrees. }
\end{figure}

\begin{figure}
\includegraphics[width=14cm,height=10cm]{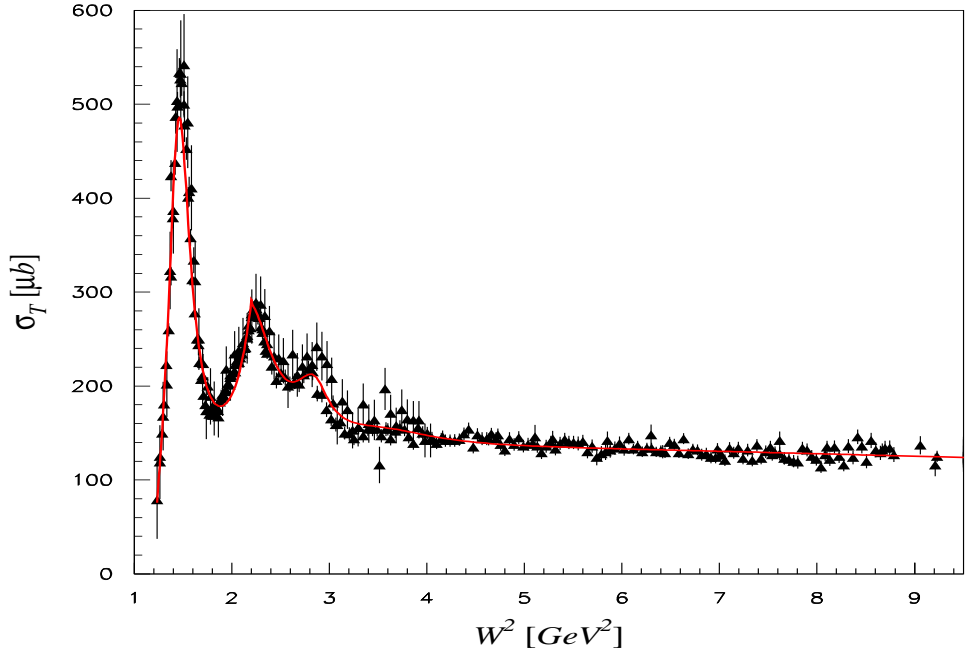}
\caption{\label{cscomp_photo} (Color online) Comparison of fit 
(solid line) to the photoproduction data used in the fit
(see Table~\ref{datasets}).}
\end{figure}

\begin{figure}
\includegraphics[width=16cm,height=20cm]{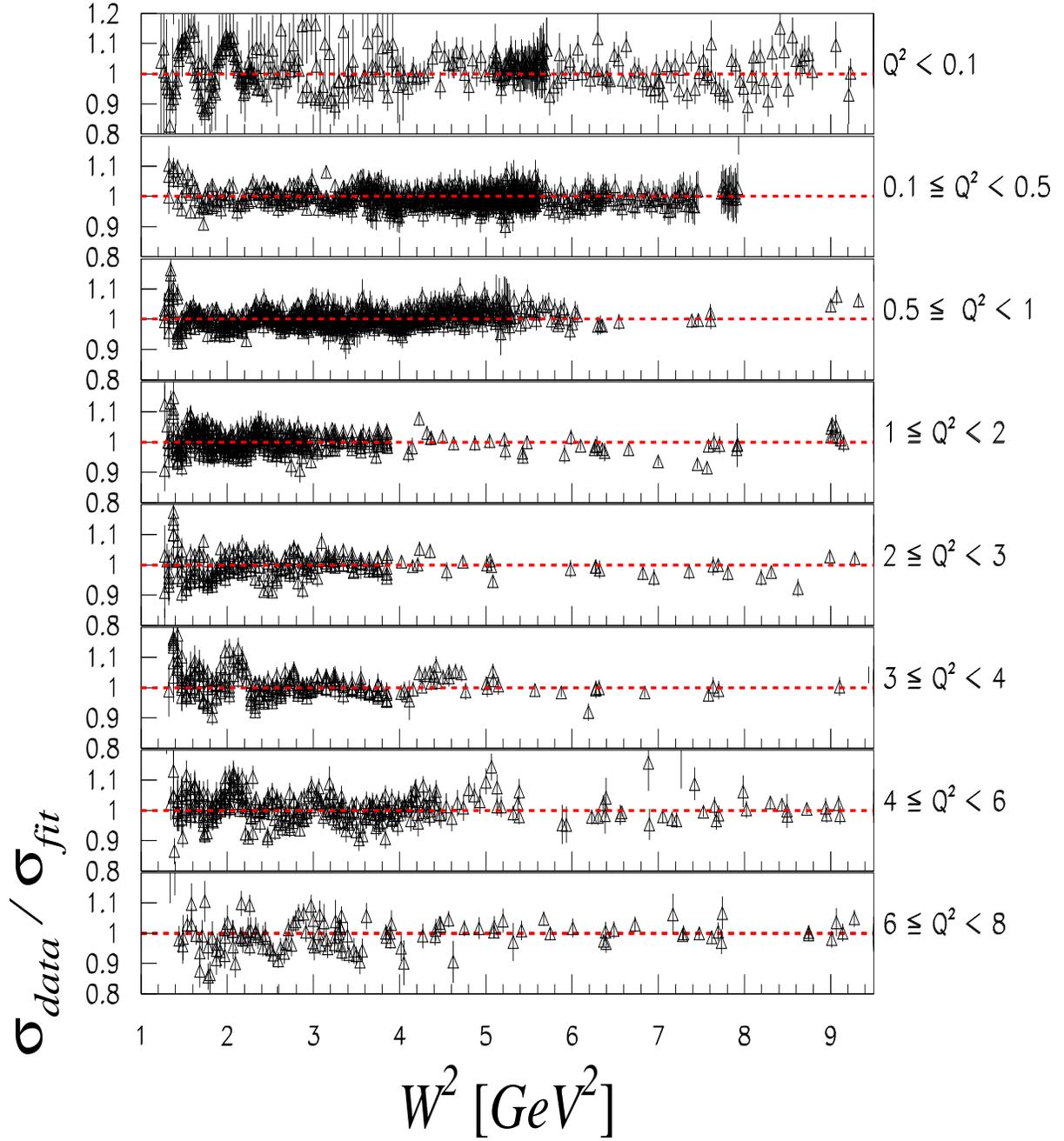}
\caption{\label{ratio_all} (Color online) Ratio of all fitted cross 
section data to the fit results at the eight $Q^2$ ranges indicated. 
The units of $Q^2$are GeV$^2$.}
\end{figure}

\begin{figure}
\includegraphics[width=16cm,height=20cm]{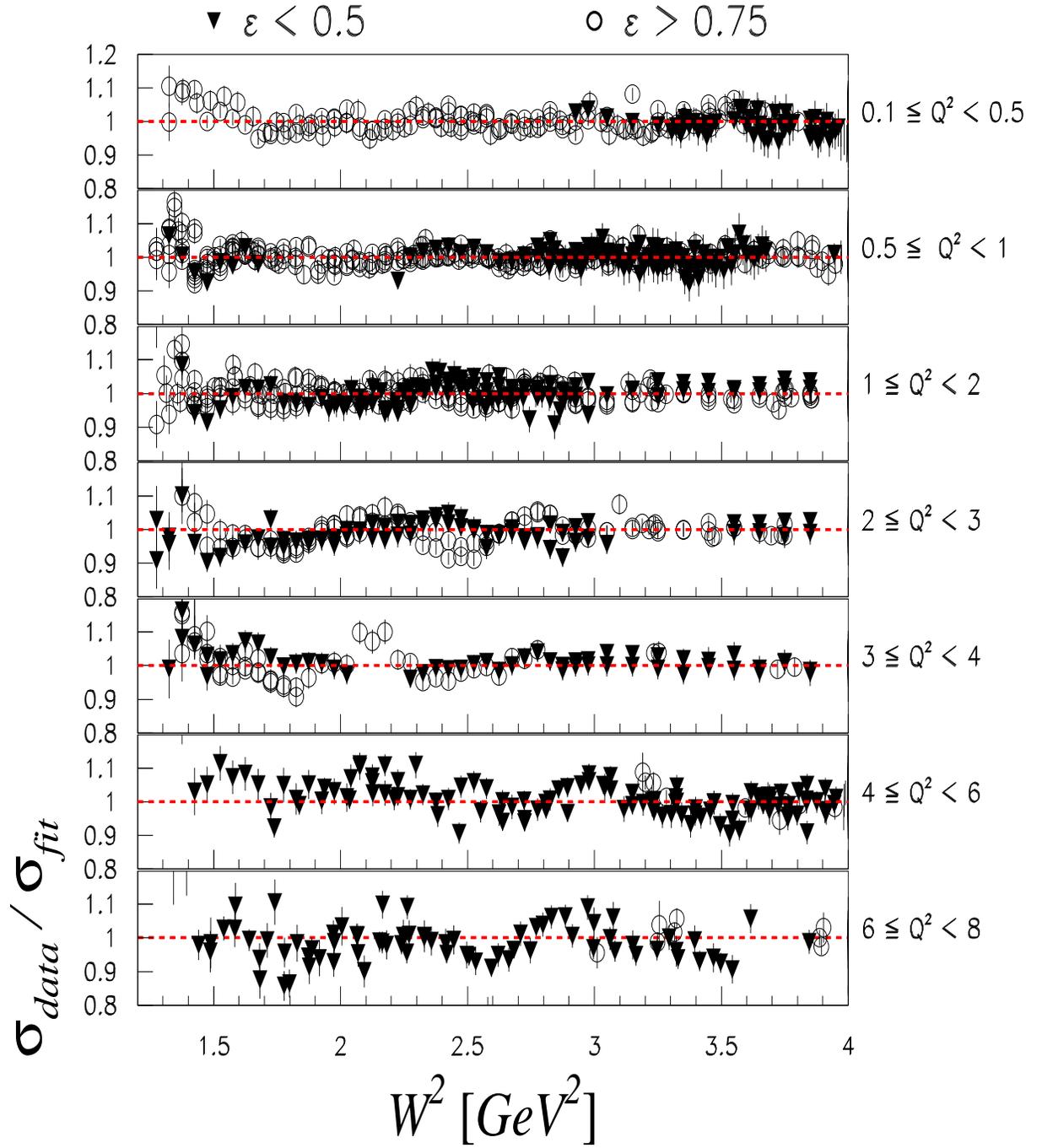}
\caption{\label{ratio_eps} (Color online) Ratio of selected 
cross section data to the fit at the seven $Q^2$ ranges 
indicated ($Q^2$ in units of GeV$^2$).  The data for $\epsilon$~$<$~0.5 are
shown as the solid triangles, while the data with 
$\epsilon$~$>$~0.75 are shown as the open circles.}
\end{figure}

\begin{figure}
\includegraphics[width=16cm,height=12cm]{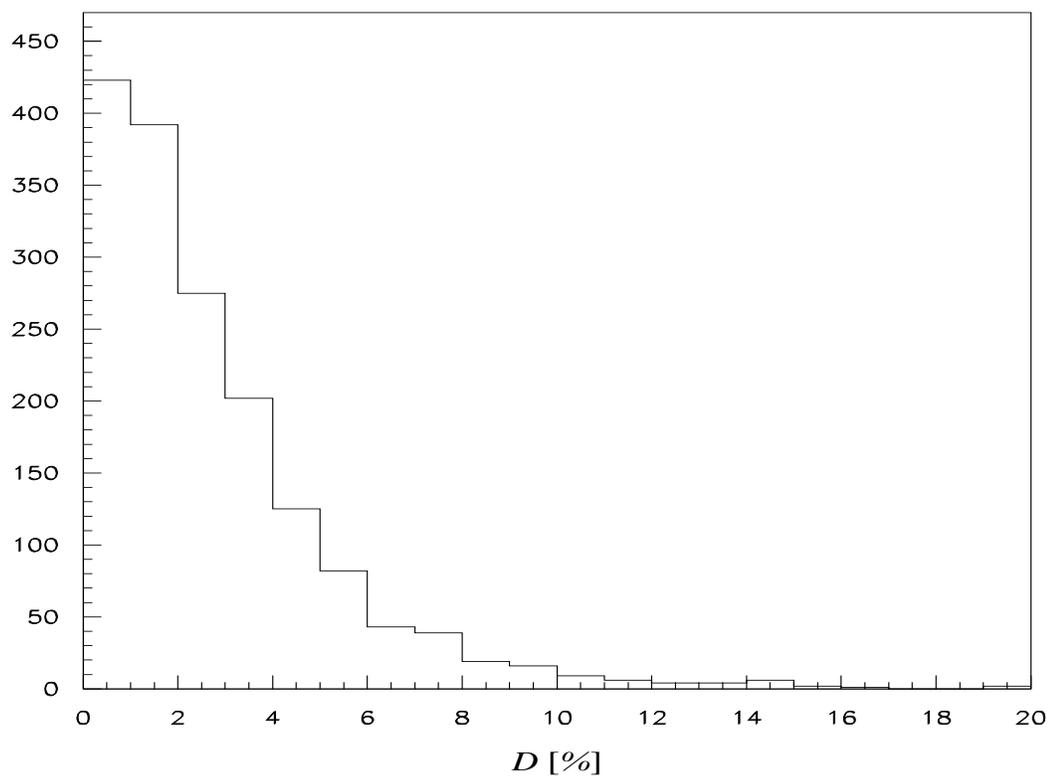}
\caption{\label{freq_pub} Frequency distribution for the percent differences of the data to the fit for 
$Q^2 > 0$.}
\end{figure}

\def\>{\hskip .15in}
\begin{table}[tbh]
\begin{center}
\begin{tabular}{r r r r r r r r r r}
\hline
$I$ & $M_i$ & $\Gamma_i$ & $A_T^i(0)$ & $a_i$ & $b_i$ & $c_i$  & $A_L^i(0)$ & $d_i$ & $e_i$ \\ 
\hline
1 &\>  1.230 &\>  0.136 &\>    7.780 &\>      4.229 &\>      1.260 &\>      2.124 &\>  29.4140 &\>     19.910 &\>    0.226\\
2 &  1.530 &  0.220 &    6.335 &   6823.200 &  33521.000 &      2.569 &   0.0 &       -    &     -   \\
3 &  1.506 &  0.083 &    0.603 &     21.240 &      0.056 &      2.489 & 157.9200 &     97.046 &    0.310\\
4 &  1.698 &  0.096 &    2.330 &     -0.288 &      0.186 &      0.064 &   4.2160 &      0.038 &    1.218\\
5 &  1.665 &  0.109 &    1.979 &     -0.562 &      0.390 &      0.549 &  13.7640 &      0.314 &    3.000\\
6 &  1.433 &  0.379 &    0.0225 &    462.130 &      0.192 &      1.914 &   5.5124 &      0.054 &    1.309\\
7 &  1.934 &  0.380 &    3.419 &      0.000 &      0.000 &      1.000 &  11.0000 &      1.895 &    0.514\\

\hline
\hline
\end{tabular}
\caption{Fit parameters for each resonance $I$, as defined in the text. Units of
cross section are $\mu b$ and all masses, momenta, and energies are in units of GeV.  
In addition to these parameters, the delta damping factor $X_1$ was determined from 
the fit.}
\label{parm_values}
\end{center}
\end{table}

\begin{table}[tbh]
\begin{center}
\begin{tabular}{r r r r r r r r r}
\hline
 L/T  &   $i$ & $\sigma_{L/T}^{NR,i}(0)$ & $a_i$ & $b_i$ & $c_i$ & $d_i$ & $e_i$ \\ 
\hline
T &\> 1 &\>   246.1 &\>  0.0675 &\>  1.3501 &\>  0.1205 &\> -0.0038 &\> -  \\
T & 2 &   -89.4 &  0.2098 &  1.5715 &  0.0907 &  0.0104 & -  \\
L & 1 &    86.7 &  0.0000 &  4.0294 &  3.1285 &  0.3340 &  4.9623 \\
\hline
\hline
\end{tabular}
\caption{Values for the transverse (T) and longitudinal (L) non-resonant fit parameters.
The units of cross section are $\mu b$ and all masses, momenta, and energies are in units 
of GeV.  In addition to those listed, the parameters $Q_0$ and $m_o$ which determine the 
transition of $\sigma_T^{NR}$ to $Q^2=0$ were determined from the fit.}
\label{nr_parm_values}
\end{center}
\end{table}

The results for the fit parameters are given in 
Tables~\ref{parm_values} and \ref{nr_parm_values}.
Overall, the fit presented here provides an excellent
description of the resonant structures seen in
inclusive $ep$ cross sections, as illustrated in Fig.~\ref{cscomp_res} 
for representative kinematic settings of the E94-110 experiment~\cite{e94110}.  
The fit does a reasonable
job in describing the photoproduction 
data~\cite{photo_bloom,photo_armstrong,photo_meyer,daphne-p}
(see Fig.~\ref{cscomp_photo}), although it was
not possible to obtain a perfect description of
the dip region between the $\Delta(1232)$ and the
second resonance region. This is illustrated more
clearly in Fig.~\ref{ratio_all}, where two
significant oscillations around unity are observed
at low $W$ and $Q^2<0.1$ GeV$^2$ (dominated
by the photoproduction data).  This is likely due to the 
$Q^2$ dependence for the transverse transition 
form factors chosen. It is certainly true that the individual 
transverse transition form factors in the second resonance 
region are {\it not} consistent with those extracted from 
exclusive analysis~\cite{burkert-lee}, although 
there is consistency in the overall 
transverse resonance strength in this region.  Part of this inconsistency 
could be caused by the chosen form for the Roper amplitude $A_T^2$.  
Recent data~\cite{burkert-lee} indicate that this amplitude decreases to zero 
by $Q^2\approx 0.4$ GeV$^2$ and then increases 
again with increasing $Q^2$.  Such a 
$Q^2$ dependence is not possible with our chosen form for $A_T^2$, and this 
could force a shifting of strength between the various resonances in this 
region which is $Q^2$ dependent.  This is a topic that will be pursued in future 
fitting studies.


The quality of the fit is quite good, overall,  
with nearly all data points differing from 
the fit by less than 5\%, and more than half of
the data points deviating from the fit by less
than 3\% as evidenced by Figure~\ref{freq_pub}. As illustrated in  Fig.~\ref{ratio_eps},
the fit describes data both at low $\epsilon$ and 
high $\epsilon$ reasonably well.  This is especially 
true for $Q^2 < 2$ where the available data show little 
or no variation in the goodness of the fit for differing  
$\epsilon$ values, indicating that both the transverse and longitudinal 
cross sections are well represented here.  The data from E94-110 clearly 
indicate that the resonance structure in the longitudinal cross section becomes 
enhanced relative to the background for $Q^2$ between 2-3~GeV, and that the 
resonanance peaks prefer widths that are narrower than for the transverse cross 
section.  This indicates that the description of the data could be improved by 
allowing the resonance width parameters to be varied independently for the 
longitudinal cross section.

The largest deviation with $\epsilon$ is
in the dip region between the second and third resonance region for 
$2<Q^2<4$ GeV$^2$.  This is mainly due 
to the difficulty in fitting the $W^2$ 
dependence of the longitudinal cross section here.  This region is 
dominated by the  E94-110 data~\cite{e94110} 
which indicate significant resonance 
structure in the longitudinal channel in this $Q^2$ range, but with the 
longitudinal strength dipping very close to zero in the dip 
region~\cite{e94110}.  
This is inconsistent with a fit form for the non-resonant cross section 
which decreases monotonically as pion threshold is approached, and could 
indicate that our assumption of incoherency between the resonant and 
non-resonant scattering is starting to fail.
Due to the lack of low $\epsilon$ data at high
$Q^2$ and $W<2$ GeV, the fit for $\sigma_L$ does
rely to some extent on the extrapolation in $Q^2$
of the fit form used. 


In summary, we have developed a fit to the total inclusive
transverse and longitudinal proton cross section
that describes existing data at the 3\% or better
level over almost the entire range $0 \le Q^2<8$ GeV$^2$
and $1.1<W<3.1$ GeV, corresponding the to the
kinematic settings available at Jefferson Lab with 
a 6 GeV beam. The fit can therefore be used to
reliably evaluate radiative corrections and to
extract spin structure functions from asymmetry
measurements. The fit also provides a convenient
representation of world data that can be used
for the experimental evaluation of the high-$x$
contribution to sum rules involving integrals over
proton structure functions. 

FORTRAN computer code embodying the fit described in
this article is available by email request from the authors.  Tables 
of cross section data fit can be found in the relevant references, 
except for the preliminary data from JLab E00-002 which will be made 
available from the Hall C website at www.jlab.org/resdata.  This 
webpage will be utilized as a repository for the final cross sections, 
as well as all available resonance region cross section data and fits.

We thank V. Tvaskis for compiling the preliminary 
data table for Ref.~\cite{Edwin}.  This work was supported in part by 
research grants 0099540 and 9633750 from the National
 Science Foundation. The Southeastern Universities
 Research Association (SURA) operated the Thomas
Jefferson National Accelerator Facility for the 
United States Department of Energy under 
contract DE-AC05-84ER40150. 


\clearpage


\end{document}